# Unresolved excess accumulation of myelin-derived cholesterol contributes to scar formation after spinal cord injury

Bolin Zheng[1], Yijing He[1], Qing Zhao[1], Xu Zhu[1], Shuai Yin[1], Huiyi Yang[1], Zhaojie Wang[1,2,*], Liming Cheng[1,*]

[1]Key Laboratory of Spine and Spinal Cord Injury Repair and Regeneration, Ministry of Education, Department of Orthopedics, Tongji Hospital, School of Medicine, Tongji University, Shanghai 200092, China.

[2]School of Life Science and Technology, Tongji University, Shanghai 200092, China.

*Corresponding authors:

Liming Cheng, e-mail: limingcheng@tongji.edu.cn; Tel: 862166111283; Fax: 862156050502

Zhaojie Wang, e-mail: 2013wangzhaojie@tongji.edu.cn; Tel: 862166111283; Fax: 862156050502

**Abstract**

**Background:** Spinal cord injury triggers complex pathological cascades, resulting in destructive tissue damage and incomplete tissue repair. Scar formation is generally considered as a barrier for regeneration in central nervous system (CNS), while the intrinsic mechanism of scar-forming after spinal cord injury has not been completed deciphered.

**Methods:** We assessed cholesterol hemostasis in spinal cord lesions and injured peripheral nerves using confocal reflection microscopy and real-time PCR analyses. The involvement of the proteins, which were predicted to promote cholesterol efflux in spinal cord lesions, were assessed with Liver X receptor (LXR) agonist and Apolipoprotein E (APOE) deficiency. The role of reverse cholesterol transport (RCT) in cholesterol clearance was examined in APOE KO mice injured sciatic nerves and myelin-overloaded macrophages in vitro. Finally, we determined the consequence of excess cholesterol accumulation in CNS by transplantation of myelin into neonatal spinal cord lesions.

**Results:** We found that excess cholesterol accumulates in phagocytes and is inefficiently removed in spinal cord lesions in young-adult mice. Interestingly, we observed that excessive cholesterol also accumulates in injured peripheral nerves, but is subsequently removed by RCT. Meanwhile, preventing RCT led to macrophage accumulation and fibrosis in injured peripheral nerves.

Furthermore, the neonatal mouse spinal cord lesions are devoid of myelin-derived lipids, and able to heal without excess cholesterol accumulation. We found that transplantation of myelin into neonatal lesions disrupts healing with excessive cholesterol accumulation, persistent macrophage activation and fibrosis, indicating myelin-derived cholesterol plays a critical role in impaired wound healing.

**Conclusion:** Altogether, our data suggest that the central nervous system (CNS) lacks an efficient approach for cholesterol clearance, resulting in excessive accumulation of myelin-derived cholesterol, thereby inducing scar-forming after injury.

**Introduction**

Spinal cord injury triggers complex pathological cascades[1], which result in destructive tissue damage, and culminate in incomplete tissue repair characterized by the formation of a scar[2, 3]. Generally, the spinal cord scar consists of a glial scar border, also known as the astrocyte scar, and a non-neural lesion core, also known as the fibrotic scar[2]. Although astrocyte scars have been proposed as the barrier to axonal regrowth, recent evidence indicated that astrocyte scar formation can beneficial for tissue repair and axon regeneration[4-6]. Whereas the fibrotic scar, primarily composed of fibroblasts and bone marrow-derived macrophages (BMDMs)[7], impedes tissue regeneration[8-10]. Notably, different from the maintained accumulation of activated macrophages/microglia in the adult lesions[7, 11], transient activation of microglia mediates glia bridge formation and organizes scar-free healing in neonatal mice[12], similar to zebrafish and newts[13, 14]. While the role of spinal cord scar is well demonstrated[10], less is known about the intrinsic mechanisms that lead to scar formation.

Cholesterol is an indispensable constituent of mammalian biological membranes. However, excess cholesterol accumulation induces ubiquitous toxicity involved in the progression of numerous diseases[15], including atherosclerosis, non-alcoholic fatty liver disease (NAFLD), chronic kidney disease (CKD), diabetes, immune dysfunction, COVID-19 and Alzheimer's disease[15]. Several mechanisms can protect cells from excess accumulation of cholesterol[16]. Preferentially, excess free cholesterol is esterified into cholesteryl esters for storage in lipid droplets. Meanwhile, cellular cholesterol is exported and delivered through reverse cholesterol transport (RCT) from peripheral tissues to the liver [16]. Under certain conditions, ineffective cholesterol efflux results in foam cell formation loaded with esterified cholesterol, free cholesterol, and crystallized cholesterol because of the hydrophobic property[17, 18].

The CNS contains more than 20% of whole-body cholesterol, up to 70% of which resides in the myelin[19]. Following a spinal cord injury, a large amount of cellular and myelin debris is generated and largely engulfed by BMDMs[20]. Although it is well known that myelin debris not only inhibits axon regeneration but also mediates inflammation[21, 22], the process and consequence of myelin degradation in phagocytes remain elusive. After spinal cord injury, abundant lipid-laden macrophages are present in the lesion core[20]. Moreover, the transcriptional profiles of these lipid-laden macrophages closely resemble foam cells at 7 days post injury (dpi), being predominantly enriched in lipid catabolism in GO analysis [23]. These findings indicate that lipids, breakdown products of cellular and myelin debris, are involved in the pathological process of spinal cord injury. Considering that a large proportion of cholesterol resides in CNS and excess cholesterol plays a crucial role in multiple diseases, it is essential to elucidate the cholesterol metabolism and its potential consequence in CNS lesions.

This study highlights the role of cholesterol homeostasis in spinal cord lesions. Using confocal reflection microscopy, we detected cholesterol crystals in spinal cord lesions as early as 7 dpi, which mediate activation of the inflammasome. We found that excess cholesterol also accumulates in injured peripheral nerves, but is subsequently removed by RCT. Moreover, preventing cholesterol efflux after peripheral nerve injury resulted in macrophage accumulation and fibrosis. We further demonstrated that myelin-derived cholesterol is necessary and sufficient for excess cholesterol accumulation in spinal cord lesions, leading to persistent macrophage activation and fibrosis.

**Results**

**1. Cholesterol crystals appear in young-adult spinal cord lesions.**

To investigate the specific role of lipids in pathological processes after spinal cord injury, we focused on the homeostasis of cholesterol since it is extremely rich in CNS. Using confocal reflection microscopy[17], we detected a large number of cholesterol crystals, a typical hallmark of excess cholesterol accumulation, as early as 7 dpi in spinal cord lesions in young-adult mice (Fig. 1a). And cholesterol crystals increased at 2 weeks post injury (wpi) and accumulated at the lesions at least for 6 weeks (Fig. 1a, f). Combined with confocal fluorescence microscopy, we detected crystals deposited in IBA1 positive phagocytes at 7 dpi (Fig. 1c, c1), but not at 3 dpi when BMDMs initially infiltrated into the lesion. Along with macrophages/microglia centripetal migration, the scar was sealed and matured at 2 wpi[24, 25] (Fig. 1d). In the lesion core of mature scar, crystals were primarily present in macrophages (Fig. 1d1). At the lesion border of the mature scar, crystals existed in both macrophages and astrocytes (Fig. 1d1, d3). Besides, crystals deposited phagocytes were also strongly positive for MAC2 (Fig. 1c2, d2), suggesting they are BMDMs[20]. Using transmission electron microscopy (TEM), the appearance of cholesterol crystals was further verified as needle-like crystals were observed in both the lysosome and the cytoplasm of foamy macrophages (Fig. 1e). In addition, tiny crystals also appear in rat spinal cord lesions (Supplementary Fig. 1). Our results suggest that phagocytes, especially macrophages, are challenged with excess cholesterol and accumulate in spinal cord lesions.

Subsequently, we quantified the time-course expression of cholesterol metabolism-related genes after spinal cord injury using real-time PCR (Fig. 1b). On the one hand, injuries resulted in reduced expression of the genes related to cholesterol synthesis. On the other hand, the injured spinal cord increased expression of the genes involved in cholesterol uptake, including *Cd36*, *Trem2* and *Apobr*, and cholesterol efflux, notably *Abca1* and *Apoe*. Meanwhile, the expression of *Acat1*, mediating cholesterol esterification, and *Ch25h*, mediating cholesterol oxidation, were also increased. Our results indicate that injured tissue is processing overload cholesterol derived from tissue debris in response to tissue damage.

**2. Excess cholesterol accumulation is associated with activation of NLRP3 inflammasome in spinal cord lesions.**

Crystal deposits induce various clinical disorders with shared molecular pathologic mechanisms, among which NLRP3 inflammasome activation is well known to trigger inflammation[26]. In atherogenesis, the formation of cholesterol crystals drives NLRP3 inflammasome-mediated inflammation by inducing lysosomal damage[17]. Moreover, *in vitro* phagocytosis of myelin debris leads to cholesterol crystallization and NLRP3 inflammasome activation[18]. Based on these findings, we examined the inflammasome-mediated caspase-1 activation, which promotes secretion of the pro-inflammatory cytokine interleukin 1β (IL-1β), in spinal cord lesions. At 7 dpi and 2 wpi, cholesterol deposited macrophages showed colocalization of LAMP1, a marker of lysosomes, and MAC2/Galectin-3, a marker of lysosomal membrane permeabilization[27] (Fig. 2a), suggesting lysosome membranes are ruptured. In addition, cathepsin D (CTSD), a lysosome protease, increased obviously from 7 dpi. Moreover, both cleaved caspase-1 and mature IL-1β were elevated persistently from 7 dpi (Fig. 2b), indicating activation of the NLRP3 inflammasome, consistent with the appearance of cholesterol crystals. Nevertheless, our results suggest excess cholesterol deposition is associated with lysosome membrane rupture and late-onset NLRP3 inflammasome activation.

Intriguingly, a white-floating layer appeared on the supernatants of centrifuged spinal cord lysates (homogenized in RIPA buffer) from 7 dpi to 6 wpi, while the layer was absent in the supernatants of lysates from sham and 3 dpi (Supplementary Fig. 2a). However, the white layer was not obvious to distinguish. To visualize the white substance, the supernatants were collected and vortexed for imaging, showing that they were opaque (identified by clarity of the word behind the tube) from 7 dpi to 6 wpi, similar to the supernatant of adipose tissue (Supplementary Fig. 2b). This pathological phenomenon indicates sustained high content of non-membrane lipids in the spinal cord lesions from 7 dpi, and it is described for the first time as we can know.

**3. Unresolved excess cholesterol accumulation in spinal cord lesions.**

Cholesterol cannot be catabolized in most mammalian cells and therefore the excess cholesterol needs to be exported to extracellular high-density lipoprotein (HDL) particles. And this process is mediated by ATP-binding cassette (ABC) transporters[16]. A previous study showed that the expression of ABCA1 in macrophages was reduced after spinal cord injury using immunostaining[20]. Since Liver X receptor (LXR), a nuclear receptor, regulates the expression of genes involved in cholesterol efflux, including *Abca1*, *Abcg1* and *Apoe*[28], we examined whether the LXR agonist, GW3965, promote cholesterol efflux out of spinal cord lesions. However, oral administration of GW3965 for 4 weeks did not reduce either the amounts of cholesterol crystals or oil red O (ORO) stained lipid droplets, where esterified cholesterol was stored (Fig. 2c-f). In addition, treatment with GW3965 also failed to improve the locomotion recovery (Supplementary Fig. 3). As the transcription of *Abca1* was up-regulated from 7 dpi (Fig. 1b), we the quantified the ABCA1 protein by western blotting analysis. Our result showed that the ABCA1 protein level is substantially increased at 2 wpi and maintained for up to 6 weeks (Fig. 2c). these data suggest that macrophages do not lack the capacity of cholesterol export in spinal cord lesions, but have defects in other elements for cholesterol transportation, such as mis-regulated cholesterol carrier.

Apolipoprotein E (APOE) is regarded as the major CNS carrier of cholesterol for delivery from astrocytes to neurons. After spinal cord injury, the expression of *Apoe* was up-regulated (Fig. 1b). Therefore, we nest examined the role of APOE in cholesterol clearance in spinal cord lesions. Unexpectedly, there were no significant differences in the deposition of cholesterol crystals and lipid droplets between APOE KO and WT mice after spinal cord injury (Fig. 2c-f). In addition, the locomotion recovery was not influenced in APOE KO mice (Supplementary Fig. 3). Our results suggest APOE is not efficient in removing excess cholesterol from spinal cord lesions.

**4. Excess cholesterol accumulation is resolved by reverse cholesterol transport in peripheral nervous system (PNS).**

A peripheral nerve injury leads to Wallerian degeneration (Fig. 3a), which produces massive myelin debris distal to the injury site. But unlike lesions in the CNS, the vast majority of myelin debris in peripheral nerve injury is rapidly cleaned by both Schwann cells and macrophages [29, 30]. We wondered whether cholesterol accumulation has a role in the difference of healing outcomes of CNS and PNS injuries. We first asked whether excess cholesterol accumulation occurs after peripheral nerve injury. At 7 dpi and 2 wpi, we observed the accumulation of cholesterol crystals and lipid droplets in the sciatic nerve distal to the crush site (Fig. 3b-e). Subsequently, cholesterol crystals and lipid droplets were eliminated at 6 wpi (Fig. 3b-e). Corresponding to cholesterol crystals deposition, active caspase-1 and mature IL-1β increased firstly after injury and then decreased at 6 wpi (Fig. 4a). These results suggest that excess cholesterol does accumulate in injured peripheral nerves and is subsequently removed.

Since peripheral nerves lack blood-tissue barriers, excess cholesterol can be transported to the liver via plasma HDL by RCT. As APOE is necessary for RCT of macrophage *in vivo*[31, 32], we then analyzed the cholesterol clearance in injured nerves of APOE KO mice. In APOE KO mice, excess cholesterol deposited in injured nerves distal to the injury site and maintained for at least 6 wpi (Fig. 3b-e), suggesting limited clearance of myelin-derived cholesterol, as that was not detected in sham group of WT and APOE KO mice (Fig. 3d, e and Supplementary Fig. 4). Furthermore, compared to WT mice, both cleaved caspase-1 and mature IL-1β increased at 6 wpi in injured nerves of APOE KO mice (Fig. 4a). These results suggest that RCT is essential for efficient cholesterol clearance in injured peripheral nerves.

**5. Efficient cholesterol clearance prevents macrophage accumulation and CSPG scarring in injured peripheral nerves.**

Next, we asked what is the consequence of excess cholesterol accumulation following nerve repair in APOE KO mice? Consistent with excess cholesterol accumulation, the injured nerves distal to the crush site were enlarged at 6 wpi in APOE KO mice, but not in WT mice (Fig. 4b). In the meantime, a large amount of CD68 positive macrophages accumulated in the injured nerves of APOE KO mice, while much less was present in WT mice (Fig. 3c, f). Furthermore, fibrosis was increased as shown with CSPG staining, though not fibronectin, in APOE KO mice injured nerves (Fig. 4c, d and Supplementary Fig. 5c, d). Besides, the density of nerve fiber stained with neurofilament 200 (NF200) was significantly reduced, similar to the previous study[33], whereas higher percentage of non-neural tissue, primarily macrophages, was present in APOE KO mice injured nerves (Supplementary Fig. 5a, b). Although APOE is not essential for nerve regeneration and remyelination in APOE KO mice[33-35], our results suggest a critical role of APOE in efficient cholesterol clearance through RCT, which prevents macrophage accumulation and CSPG scarring in injured peripheral nerves.

**6. Serum is efficient for the clearance of myelin-derived cholesterol *in vitro*.**

For a better understanding of the different consequences in response to cholesterol accumulation between CNS and PNS, we performed *in vitro* experiment to study the role of serum in the clearance of myelin-derived cholesterol.

The serum contains HDL and other cholesterol-binding agents, which stimulate cholesterol secretion in cultured macrophages[36, 37]. To study the process of cholesterol efflux in response to myelin overloading, we first incubated the BMDMs with myelin debris for sufficient endocytosis of cholesterol (Fig. 5a). After incubation, BMDMs were enlarged, in which a few lipid droplets and cholesterol crystals appeared (Fig. 5b-d), suggesting that myelin debris was internalized by BMDMs and cholesterol was already partly released from degrading myelin debris. Then we cultured the myelin-loaded cells in Dulbecco's modified Eagle's medium (DMEM) supplemented with or without 10% fetal bovine serum (FBS), the washing off the free myelin debris (Fig. 5a). In the absence of FBS, a large number of lipid droplets and cholesterol crystals appeared in BMDMs 24 hours later and persisted up to 48 hours. But in the appearance of FBS, despite the existence of lipid droplets in BMDMs at 24 hours of culture, neither lipid droplets nor cholesterol crystals is detectable after 48 hours. (Fig. 5b-d). Besides, as a positive control, cholesterol treatment led to cholesterol deposition in BMDMs, which was subsequently removed by FBS (Fig. 5e, f, Supplementary Fig. 6c). As a negative control, sphingomyelin, a member of membrane lipids, had no contribution to the formation of lipid droplets and cholesterol crystals (Fig. 5e, f, Supplementary Fig. 6b). Furthermore, we cultured myelin-overloaded BMDMs in DMEM supplemented with HDL and observed that myelin-derived cholesterol was efficiently removed after 48 hours (Fig. 5g-i). Together, our results suggest myelin-derived cholesterol cannot be efficiently removed from macrophages unless an effective acceptor of cholesterol is present in the culture medium.

Previous studies indicated that myelin debris is involved in macrophage polarization[20, 38]. We next analyzed macrophage polarization and the related inflammation in response to excess cholesterol accumulation. As described above, myelin-overloaded BMDMs were cultured in different mediums without any other stimulators for 48 hours (Fig. 5a). In response to excess cholesterol accumulation, BMDMs up-regulated the expression of both *Il6* and *Inos*, related to pro-inflammatory factors of M1 macrophages, and the up-regulation was reversible with removal of accumulated cholesterol by FBS or HDL (Fig. 5j, k). But excess cholesterol accumulation did not influence the expression levels of *Tnf*, an M1 macrophage marker, and *Arg1, Tgfb, Igf1* and *Il10*, M2 macrophage markers (Supplementary Fig. 6d-h). Furthermore, we found that the expression of *Nlrp3*, *Casp1* and *Il1b* were up-regulated in response to excess cholesterol accumulation, while all of them were not up-regulated in the appearance of FBS or HDL (Fig. 5l-m), suggesting the increased expression of NLRP3 inflammasome components induced by excess cholesterol accumulation were rescued when cholesterol is removed. These results suggest that excess cholesterol accumulation induces specific pro-inflammatory responses in macrophages, but not the same as M1/M2 activation[23].

**7. myelin-derived cholesterol leads to persistent macrophage activation and scar formation in spinal cord lesions.**

Spinal cord injury produces myelin debris and other biological membrane debris. Considering that the biological membranes contain plentiful lipids, including cholesterol, we asked whether myelin-derived cholesterol was necessary for cholesterol accumulation following spinal cord injury. Using luxol fast blue (LFB) to stain myelin lipids we observed that myelin lipids are not present in the postnatal day 1-3 (P1-3) mouse spinal cord, but are abundant in the spinal cord after P7, as myelination proceeds (Fig. 6a and Supplementary Fig. 7a). We detected the expression of myelin basic protein (MBP), a myelin-related protein, in the mouse spinal cord, starting at P2 (Fig. 6b and Supplementary Fig. 7b).

Next, we performed P2 injuries for further study of the roles of myelin-derived lipids[12]. We found that at 2 weeks post P2 crush injury, spinal cord lesions are repaired, without deposition of cholesterol crystals (Fig. 6c), Next, we injected heat-inactivated (95°C, 15 minutes) myelin debris into the lesion site immediately after a P2 crush injury. As expected, we detected cholesterol crystals deposited in macrophages at the lesion core at 2 weeks post myelin debris injection (Fig. 6c, f). Meanwhile, myelin injection led to a significant accumulation of IBA1-MAC2-CD68 positive macrophages (activated macrophages) at the lesion site, and a scar was formed (Fig. 6d-i, and Supplementary Fig. 7c-e). At the same time, we observed fibrosis stained with fibronectin, chondroitin sulfate proteoglycan (CSPG) and laminin after myelin debris injection (Fig. 6j-m). These results suggest cholesterol derived from myelin debris leads to excess cholesterol accumulation and scar-forming in neonatal lesions, similar to the adult lesions.

It is reported that immune cell infiltration drives CNS fibrosis by proliferative fibroblasts[9]. To determine the role of cholesterol overloaded macrophages in scar-forming, we first overloaded BMDMs with cholesterol by incubating them with myelin debris and then transplanted myelin overloaded BMDMs into the P2 injury site. At 2 weeks post-transplantation, cholesterol overloaded BMDMs clustered closely in the lesion, surrounded by astrocytes (Fig. 6c, d, f-i, and Supplementary Fig. 7c-e). Similarly, we detected fibrosis after transplantation (Fig. 6j-m). As a control, transplantation of untreated BMDMs did not show cholesterol crystal deposition, active macrophage accumulation or fibrosis (Fig. 6c, d, f-m and Supplementary Fig. 7c-e). These results suggest that excess cholesterol accumulation leads to persistent macrophage activation, which impairs complete healing and promotes scar-forming.

Taken together, our results suggest that myelin-derived cholesterol results in not only excess cholesterol deposition but also impaired scar-free healing with macrophage activation and fibrosis in neonatal spinal cord lesions.

**Discussion**

In CNS lesions, the formation of a scar impairs spontaneous wound healing. Here, we revealed one of the intrinsic mechanisms that lead to scar-forming. Unresolved excessive myelin-derived cholesterol accumulates in spinal cord lesions, inducing persistent macrophage activation and scar formation. In comparison, efficient cholesterol clearance by RCT prevents macrophage accumulation and CSPG scarring in injured peripheral nerves.

It is reported that pharmacological stimulation can rescue cholesterol export deficiency of phagocytes both *in vitro* and *in vivo*[18, 39, 40]. However, administration of GW3965, an LXR agonist, fails to rescue excess cholesterol accumulation in spinal cord lesions in young-adult mice (Fig. 2). Our data suggest that macrophages in spinal cord lesions do not lack the cholesterol export capacity but lack other elements for cholesterol transportation.

As an *in vivo* example, in the injured peripheral nerves, excess cholesterol is efficiently removed by RCT, which is absent in the CNS. Furthermore, our *in vitro* experiment demonstrated that myelin-derived cholesterol could not be efficiently removed from macrophages unless an effective cholesterol acceptor is present (Fig. 5). Beyond the perspective of this study, it is reported that adiponectin reduces myelin lipid accumulation after spinal cord injury[41]. In addition, deficiency of cholesterol import-related receptors partly reduces the formation of foamy macrophages [23, 42]. Thus, our study demonstrates that developing an efficient strategy for excess cholesterol removal or recycling may rescue impaired healing in CNS lesions.

Although the role of APOE in CNS disease is widely explored, the potential influence of APOE on peripheral nerve injury remains incompletely understood[33-35, 43-48]. Here we unveiled the role of APOE in cholesterol clearance after peripheral nerve injury. APOE plays a role in RCT[31, 32] that efficiently removes myelin-derived cholesterol produced by Wallerian degeneration. Remarkably, excess cholesterol accumulation relates to macrophage accumulation and CSPG deposition in the peripheral nerve of APOE KO mice (Fig. 3 and Fig. 4). However, in the CNS, APOE loss-of-function does not lead to aggravated cholesterol accumulation (Fig. 2). This discrepancy may due to the much larger lesion (more than ten folds) in our study than the demyelinating lesion induced by a lysolecithin injection, which implicated a requirement for APOE[18].

This study shows that formation of cholesterol crystals at the CNS lesion sites induces lysosomal membrane permeabilization and inflammasome activation. Moreover, the increased expression of NLRP3 inflammasome components induced by excess cholesterol accumulation were rescued when cholesterol is removed in the appearance of FBS or HDL *in vitro*. Other cholesterol-induced toxicities may also participate in the pathological process of CNS injury, such as cholesterol-induced inflammation, endoplasmic reticulum stress, oxidative stress and mitochondrial dysfunction[15, 49, 50]. In addition to the macrophages we show in this study, microglia, astrocytes and even microvascular endothelial cells may also be challenged by the accumulation of excess cholesterol derived from myelin[51]. Nevertheless, our study provides a foundation to further explore the consequence of excess cholesterol accumulation in the CNS.

We performed transplantation studies to examine the role of myelin-derived lipids in spinal cord lesions. The neonatal (P2) spinal cord is absent of myelin-derived lipids, and can be completely repaired without scar formation after injury[12]. This provides us an ingenious model to test the role of myelin-derived lipids. We showed that similar to adult spinal lesions[20], cholesterol-overloaded macrophages persist in neonatal lesions after transplantation of myelin or myelin-overloaded BMDMs, leading to formation of a scar instead of healing (Fig. 6). Consistently, reduced cholesterol clearance in APOE KO mice inhibit healing of the peripheral nerves (Fig. 3, and Supplementary Fig. 5). Notably, in the P2 neonatal injury, transient activation of BMDMs and microglia is induced at 3 dpi but are eliminated by 2 wpi[12]. Though the underlying mechanism of persistent macrophage activation mediated by excess cholesterol remains to be uncovered, our results suggest excess cholesterol accumulation contributes to scar-forming mediated by persistent macrophage activation.

**Conclusion**

Our study highlights that excess myelin-derived cholesterol accumulation contributes to persistent activation of macrophages and scar-forming after spinal cord injury. Since efficient cholesterol clearance by RCT prevents macrophage accumulation and fibrosis in injured peripheral nerves, this study suggests that promoting cholesterol clearance and reestablishing homeostasis of cholesterol-overloaded macrophages are potential strategies to facilitate scarless healing in the CNS.

**Abbreviations**

**RCT:** Reverse cholesterol transport

**CNS:** Central nervous system

**PNS:** Peripheral nervous system

**BMDMs:** Bone marrow-derived macrophages

**HDL:** High density lipoprotein

**dpi:** Days post injury

**wpi:** Weeks post injury

**P2:** Postnatal day 2

**ORO:** Oil red O

**TEM:** Transmission electron microscopy

**CTSD:** Cathepsin D

**ABCA1:** ATP-binding cassette transporter A1

**APOE:** Apolipoprotein E

**MBP:** Myelin basic protein

**CSPG:** Chondroitin sulfate proteoglycan

**NF200:** Neurofilament 200

**DMEM:** Dulbecco's modified Eagle's medium

**WT:** Wild type

**Methods:**

**Animals**

All animal experiments were performed in compliance with protocols approved by the Institutional Animal Care and Use Committees (IACUC) of Tongji University. All mice were housing under temperature-controlled conditions on a 12-hour light-dark cycle, with ad libitum feeding. C57BL/6 mice and APOE KO mice purchased from Shanghai Jiesijie Laboratory were used in this study. APOE KO mice were maintained on a C57BL/6 genetic background. And genotype identification was performed with the following primers: common forward GCC TAG CCG AGG GAG AGC CG; wild type reverse TGT GAC TTG GGA GCT CTG CAG C; mutant reverse GCC GCC CCG ACT GCA TCT.

**Spinal cord injury**

A spinal cord contusion injury was conducted in 8-12-week-old female mice at thoracic level 10 (T9). Briefly, mice were anesthetized with isoflurane (2-3%), the thoracic vertebras were exposed, followed by a T9 laminectomy. Mice were then stabilized, and a moderate contusion was performed using an NYU/MASCIS impactor from the height of 12.5 mm. The muscles and skin were sutured. Bladders were manually expressed 2 times per day.

A neonatal crush injury was conducted in P2 mice at T10. The spinal cord was exposed entirely and was then crushed for 3 seconds using a Dumont No. 5 forceps. The muscles and skin were sutured. Neonatal mice were then warmed up and rubbed with the urine from the mom at the skin wound. PBS (0.5 μl), heat-inactivated myelin debris (0.5 μl, protein concentration: 4.23 μg/μl),

BMDMs or myelin-overloaded BMDMs ($1\times10^5$) were immediately injected into the lesion site after a crush injury.

**Sciatic nerve crush injury**

Sciatic nerve injuries were conducted in 8-12-week-old female mice. Mice were anesthetized with isoflurane (2-3%). Following the shaving of the hind limb area, an incision was made at the Left thigh, and sciatic nerves were gently exposed and crushed for 30 seconds using a Dumont No. 5 forceps (at 1.5 mm from the tip). After crushing, the crush site should be translucent. And then the wound was closed by sutures.

**Drug administration**

For LXR agonist administration, Mice were fed with powdered standard rodent chow added with GW3965 (20 mg/kg), according to a previous study[18]. The administration was started immediately after injury and continued for 4 weeks.

**Behavioral evaluations**

We evaluated locomotor recovery when mouse was walking in an open field using the Basso mouse scale (BMS) scores at 3 days after spinal cord injury, and weekly thereafter[52]. The mouse was evaluated without any stimulation for 5 minutes after environment adaptation for 2 minutes. The evaluation was performed by two experimenters in a blinder way. 8 mice per group was evaluated.

**Immunohistochemistry and imaging**

After anesthesia by overdose of sodium pentobarbital, mice were transcardially perfused with PBS followed by 4% paraformaldehyde. Spinal cords and sciatic nerves were dissected and fixated further with immersion in 4% paraformaldehyde at 4°C overnight and subsequently dehydrated in buffered 30% sucrose for 24 hours. Equilibrated tissues were embedded in OCT for longitudinal sectioning and stored at -80°C until processing. 20 μm thick frozen sections were prepared using the Leica CM3050S cryostat microtome. Longitudinal sections of spinal cord were numbered in order from the ventral to the dorsal side, and sections of the same/adjacent number from different mice were used for each staining.

For immunohistochemistry analysis, sections were permeabilized and blocked with blocking buffer (0.1% triton, 0.05% tween-20 and 10% normal donkey/goat serum in PBS) for 1 hour at room temperature before primary antibody incubation. The following primary antibodies were used: rat anti-CD68 (1:1000, ab53444, Abcam), rabbit abti-IBA1 (1:1000, 019-19741, Wako), rabbit anti-GFAP (1:1000, Z0334, Dako), goat anti-GFAP (1:1000, ab53554, Abcam), rat anti-MAC2 (1:500, 125401, BioLegend), rabbit anti-LAMP1 (1:200, bs-1970R, Bioss), rat anti-CD107a (1:1000, CL488-65050, Proteintech), rabbit anti-Fibronectin (1:200, F3648, Sigma), rabbit anti-Laminin (1:200, L9393, Sigma), mouse anti-Chondroitin sulfate (1:200, c8035, Sigma), rat anti-MBP (1:200, MAB386, Sigma). Secondary antibodies: Alexa Fluor 488, 555-conjuated donkey anti goat/rabbit antibodies (1:500, Invitrogen); Alexa Fluor 488, 594-conjuated goat anti rat/rabbit/mouse antibodies (1:500, Invitrogen). Sections were counterstained with DAPI and mounted with Fluoromount-G anti-fade mounting medium (Southern Biotechnology). Tissue sections were imaged using a confocal laser scanning microscope (LSM 880, Zeiss), and sections of each experimental group were photographed using consistent exposure settings. For MBP staining, the sections were pre-treated with 95% ethanol for 10 minutes.

**Crystal imaging**

Confocal reflection microscopy, which detects backscattered light to create high-resolution images, was used to visualize cholesterol crystals as described previously[17]. For confocal reflection imaging, frozen sections were scanned with a 408 nm laser, and the reflected light was collected. Confocal reflection and fluorescence microscopy (LSM 880, Zeiss) were combined to scan multiply-labeled tissues.

**Transmission electron microscopy**

At 2 wpi, the spinal cord was isolated after perfusing with 4% paraformaldehyde and fixed in 2.5% glutaraldehyde. The 1 mm tissue of the lesion center was then dissected, postfixed in 1% osmium tetroxide, dehydrated through a series of graded alcohols, cleared in acetone and embedded in Epon. Ultrathin sections (60 nm) were made using an ultramicrotome, collected on copper grids, and stained with uranyl acetate and lead citrate. After drying, the grids were viewed on a 120 kV transmission electron microscopy (HITACHI HT 7800).

**Oil Red O staining**

To visualize neutral lipids, primarily triacylglycerol and cholesterol esters, frozen sections were rinsed with 60% isopropanol for 30 seconds, then proceeded to freshly prepared Oil Red O working solution (G1261, Solarbio) for 15 minutes. Next, slides were differentiated in 60% isopropanol, rinsed with distilled water, and mounted for further imaging.

**Luxol fast blue staining**

Frozen sections were incubated in luxol fast blue solution (G3245, Solarbio) for 24 hours at room temperature. Sections were then rinsed in water, differentiated in lithium carbonate solution for 15 seconds, and differentiated further in 70% alcohol reagent for 30 seconds. After that, sections were rinsed in water and mounted.

**RNA isolation and real-time PCR**

For expression analysis, a 3 mm spinal cord of the lesion center was dissected and homogenized in RNAiso Plus (9108, Takara) using a Cryogenic grinder （Shanghai Jingxin JXFSTPRP-CL). According to the manufacturer's manual, total RNA was isolated, followed by measurement of the RNA concentration and quality using a NanoDrop spectrophotometer (Thermo Scientific). After cDNA synthesis with a PrimeScript™ RT Master Mix (RR036A, Takara), real-time PCR was performed using a TB Green® Premix Ex Taq™ II (RR820A, Takara) on a QuantStudio 7 Flex real-time PCR system (Thermo Scientific). All primers were listed in Supplementary Table 1. Using the ΔΔ-Ct-method, the expression level of target genes was calculated (β-actin as a housekeeping gene, sham group as reference samples).

**Western blot**

Spinal cords, 3 mm of the lesion center, and sciatic nerves, 5 mm distal to the crush site, were extracted and homogenized in RIPA lysis buffer (P0013B, Beyotime) supplemented with protease and phosphatase inhibitor cocktail and 5mM EDTA (P1051, Beyotime). The supernatants of tissue lysates were collected after centrifugation (12,000 g for 20 minutes at 4°C) and protein concentrations were measured using a protein BCA Protein Assay Kit (KGP903, Keygen). 20 μg of total protein were separated using the SDS-PAGE gel and transferred to PVDF membranes. The membranes were blocked with 5% milk for 1 hour at room temperature and then incubated with primary antibody overnight at 4°C. Following primary antibodies were used: rabbit anti-ABCA1 (1:1000, bs-23418R, Bioss), rabbit anti-caspase-1 (1:1000, 22915-1-AP, Proteintech), rabbit anti-IL-1β (1:1000, 16806-1-AP, Proteintech), rabbit anti cathepsin D (1:1000, 21327-1-AP, Proteintech), rabbit anti-β-actin (1:1000, 8457, Cell signaling technology). After washing in TBST, the membranes were incubated with HRP–conjugated goat anti-rabbit IgG (1:3000, ab6721, Abcam) at room temperature for 1 hour. The membranes were visualized by chemiluminescence (WBKLS0500, Millipore).

**Myelin isolation**

Myelin was extracted from adult C57BL/6 mouse brains using sucrose-gradient centrifugation as previously described[53], with some modifications[54]. At the last step, the purified myelin fraction was resuspended in PBS and stored at −80 °C. Total protein concentration was measured using a protein BCA Protein Assay Kit (KGP903, Keygen)

**BMDM isolation and culture**

To prepare bone marrow-derived macrophages (BMDM), 8-week-old mice were sacrificed by cervical dislocation and soaked with 75% ethanol. Then the femur and tibia were exposed, dissected and flushed with DMEM using a 2-ml syringe. The bone marrow cells were then filtered through a 100 μm cell strainer and cultured in basal medium (DMEM supplemented with 1% GlutaMAX and 1% penicillin-streptomycin) supplemented with 20 ng/ml M-CSF and 10% FBS for at least 7 days. For endocytosis, mature BMDMs were cultured in basal medium supplemented with 10% FBS and incubated with myelin debris (100 μg myelin protein per milliliter), cholesterol (100 μg/ml) and sphingomyelin (100 μg/ml), respectively, for 6 hours. Mediums were sonicated for 15 minutes before adding. For the study of cholesterol efflux, cells were then cultured in basal medium, basal medium + 10% FBS and basal medium + HDL (0.1 mg/ml), respectively, for additional 24 hours or 48 hours.

**Quantification**

All images were measured using Fiji software. During preparation of the longitudinal sections of tissue, each section was numbered in order, and sections of the similar position/level from different mice were selected as described above. One section in the center from each individual was used for staining and measurement, and the number of mice analyzed is described in the figure legends. For quantification of ORO, CD68, CSPG, Fibronectin and NF200 positive area (%) in sciatic nerve sections, the average value of each section was measured with 3 fields (300 μm wide squares), which were selected along the crush-distal axis from 500 μm distal to the crush site (Wallerian degeneration area), every 300 μm apart. For quantification of ORO positive area and crystal intensity of BMDMs in vitro, 3 fields from each culture (3 cultures) were measured and normalized to cell numbers. For quantification of CD68, IBA1 MAC2, Fibronectin, Laminin and CSPG in neonatal spinal cord lesions, the positive area was measured at the lesion site, and the GFAP intensity was measure at the border. The spinal cord lesion was defined as the GFAP-negative region. For quantification of crystal in tissues, the intensity of backscattered signals was measured in indicated area (lesion core or lesion border of adult spinal cord lesions, Wallerian degeneration area of injured sciatic nerves), and the intensity index was calculated by normalizing to the proximal /intact region of the spinal cord or sciatic nerve[12].

**Statistical analysis**

All data analyses were performed with GraphPad Prism. The number of samples is described in the figure legends, and data distribution was assumed to be normal. For comparison between two columns, two-tailed Student's t test was performed. For comparison between multiple columns, ordinary one-way ANOVA with turkey's multiple comparisons test was performed. For comparison between multiple columns of grouped data (Fig.3), ordinary two-way ANOVA with turkey's multiple comparisons test was performed. For comparison of the BMS scores, repeated measures two-way ANOVA with the Geisser-Greenhouse correction was performed. All values are shown as mean ± sem. *$p < 0.05$, **$p < 0.01$, ***$p < 0.001$.

**Declarations**

**Ethics approval and consent to participate**

All animal experiments were performed in compliance with protocols approved by the Institutional Animal Care and Use Committees (IACUC) of Tongji University.

**Consent for publication**

Not applicable.

**Availability of data and materials**

The data that support the findings of this study are available from the corresponding author upon reasonable request.

**Competing interests**

The authors declare that they have no competing interests.

**Funding**

This work was financially supported by the International Cooperation Project of National Natural Science Foundation of China (Grant No.81810001048), the National Natural Science Foundation of China (Grant No.81922039 and 81873994), Key basic research projects of Shanghai Science and Technology Commission (Grant No. 19JC141470), and the Fundamental Research Funds for the Central Universities of China.

**Author contributions**

L.C., B.Z., and J.W. conceptualized the project and designed experiments. B.Z., Q.Z, X.Z, S.Y, and H.Y. performed experiments. Y.H analyzed data. B.Z, and Y.H. contributed to manuscript preparation. Q.Z. provided technical or material supports.

**Acknowledgements**

We thank Gufa Lin and Margaret S. Ho for their assistance of manuscript preparation.

**Figures**

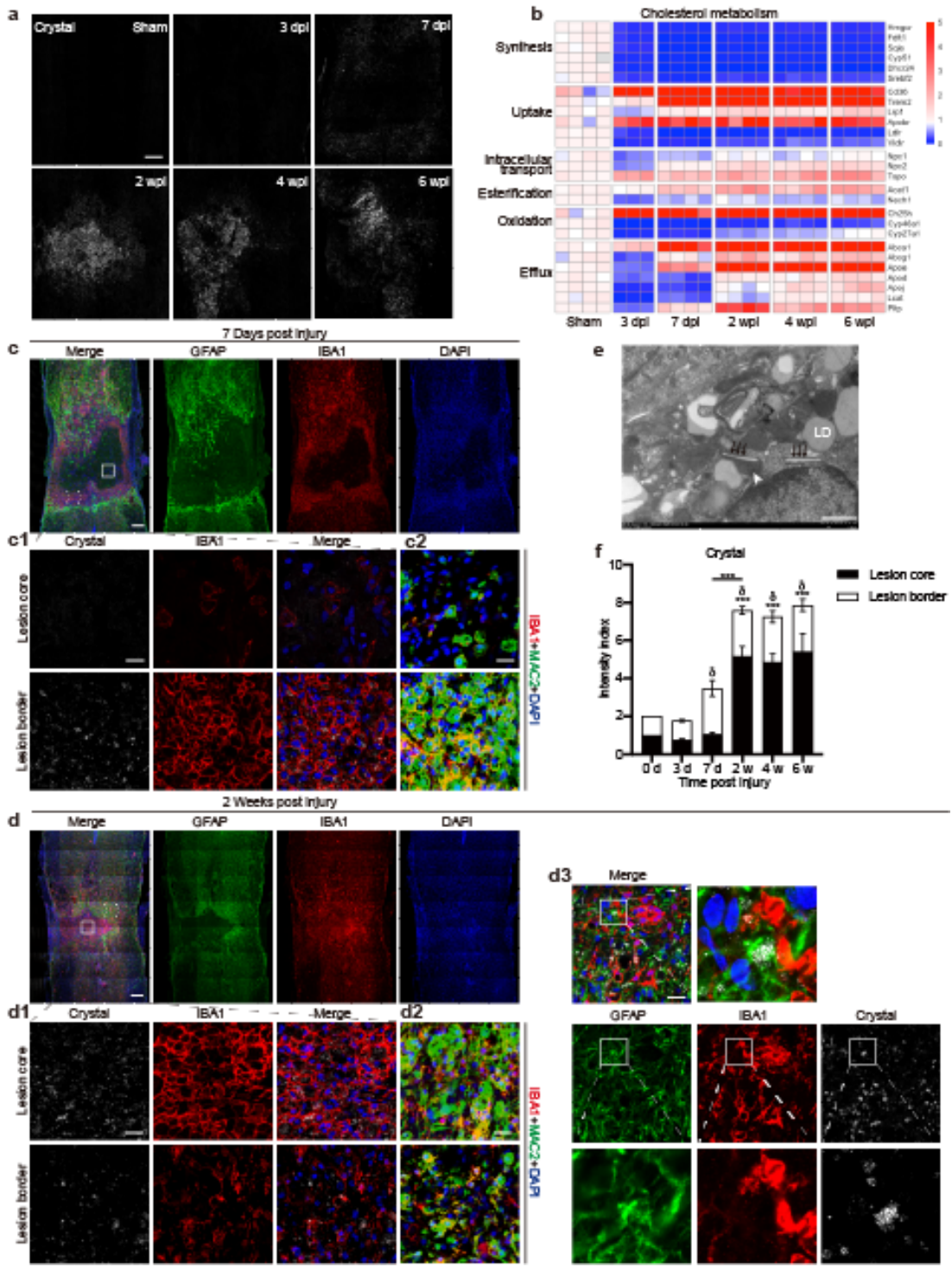

**Fig.1: Accumulation and distribution of cholesterol crystals in spinal cord lesions. a**, Representative reflection images of spinal cord lesions of young-adult mice at different time points after injury; The reflection signal is shown as white. **b**, Quantification of the time-course expression of cholesterol

metabolism-related genes in spinal cord lesions using real-time PCR. **c**, **d**, Confocal images of spinal cord lesions stained with GFAP (green), IBA1(red) and DAPI (blue) at 7 dpi and 2 wpi, respectively. **c1**, **d1**, Combined confocal fluorescence and reflection microscopy images of the boxed areas (box: lesion core, dashed box: lesion border) from **c** and **d**, respectively, showing crystals (white) in IBA1 positive phagocytes (red). **c2**, **d2**, Representative images of the lesion core and the lesion border stained with IBA1(red), MAC2(green) and DAPI (blue) at 7 dpi and 2 wpi, respectively. **d3**, Representative images of the lesion border at 2 wpi stained with GFAP (green), IBA1(red) and DAPI (blue), showing crystals (white) also in GFAP positive astrocytes (green). Boxed areas are shown magnified. **e**, TEM image of spinal cord lesion at 2 wpi, showing needle-like cholesterol crystals (black arrows) and lysosome (white arrowhead). LD: lipid droplet. **f**, Quantification of crystal intensity (normalized to the proximal region) in the lesion core and the lesion border (n=4 mice). Ordinary one-way ANOVA with turkey's multiple comparisons test. Data are shown as mean ± sem. ***$p < 0.001$ (lesion core, each column vs the column of Sham unless indicated). $^{\delta} p < 0.05$ (lesion border, each column vs the column of Sham). Scale bar: 200 μm (a, c, d), 20 μm (c1, c2, d1, d2, d3), 1 μm (e).

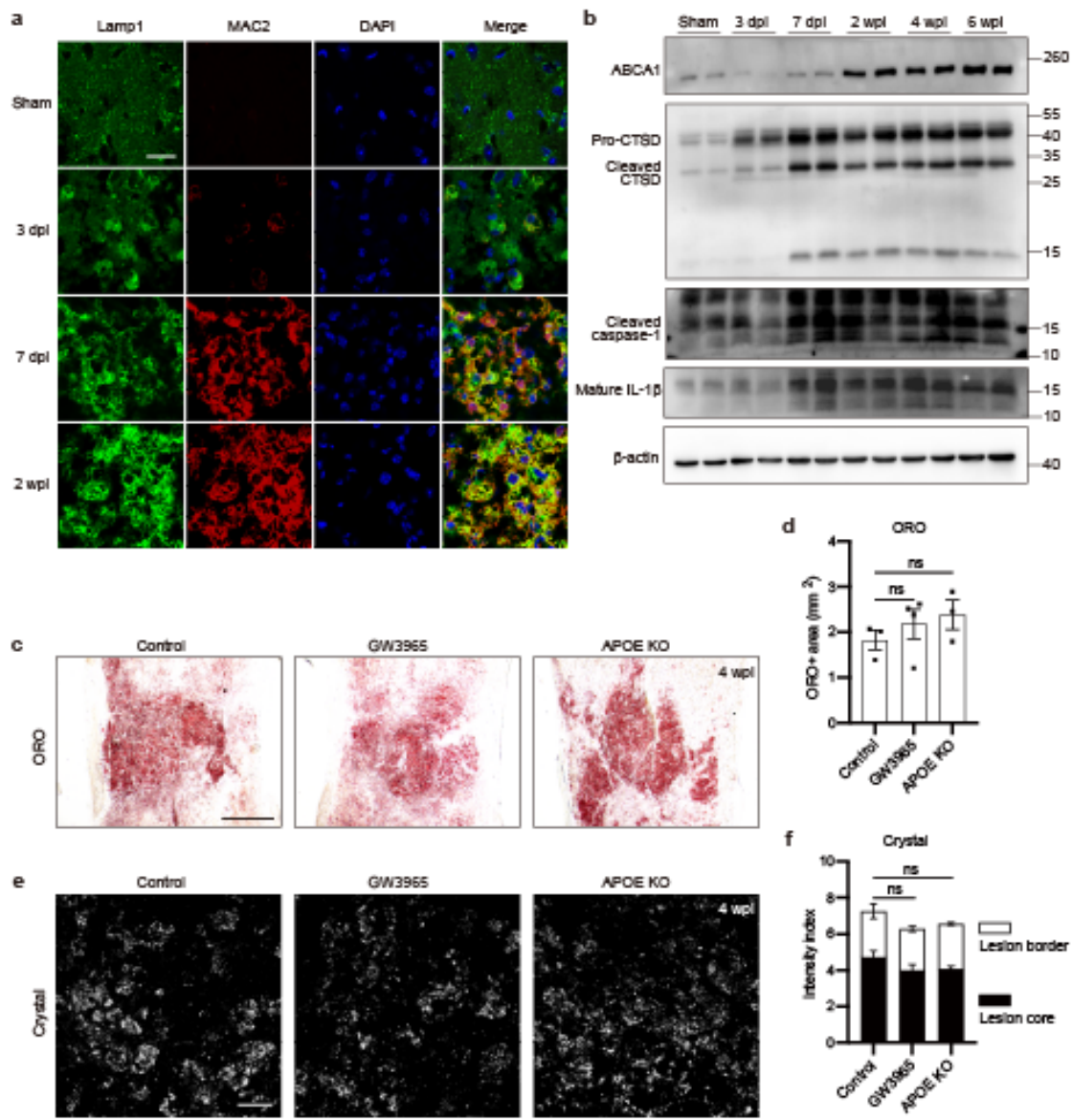

**Fig.2: Unresolved excess cholesterol deposition and inflammation in spinal cord lesions. a**, Representative images of spinal cord lesions stained with LAMP1 (green), MAC2 (red) and DAPI (blue) at different time points after injury. **b**, Immunoblot of spinal cord lesions at different time points after injury for ABCA1, CTSD, Caspase-1, IL-1β and β-actin (loading control). **c**, Representative images of spinal cord lesions from different groups of mice at 4 wpi stained with ORO (red). **d**, Quantification of ORO positive areas from d (n=3, 4 and 3 mice for Control, GW3965 and APOE KO, respectively). **e**, Reflection

images of spinal cord lesion core from different groups of mice at 4 wpi, showing crystals (white). **f**, Quantification of crystal intensity (normalized to the proximal region) from **f** (n=3 mice). **d**, **f**, Ordinary one-way ANOVA with turkey's multiple comparisons test. All data are shown as mean ± sem. ns: not significant. Scale bar: 20 µm (**a**, **e**), 200 µm (**c**).

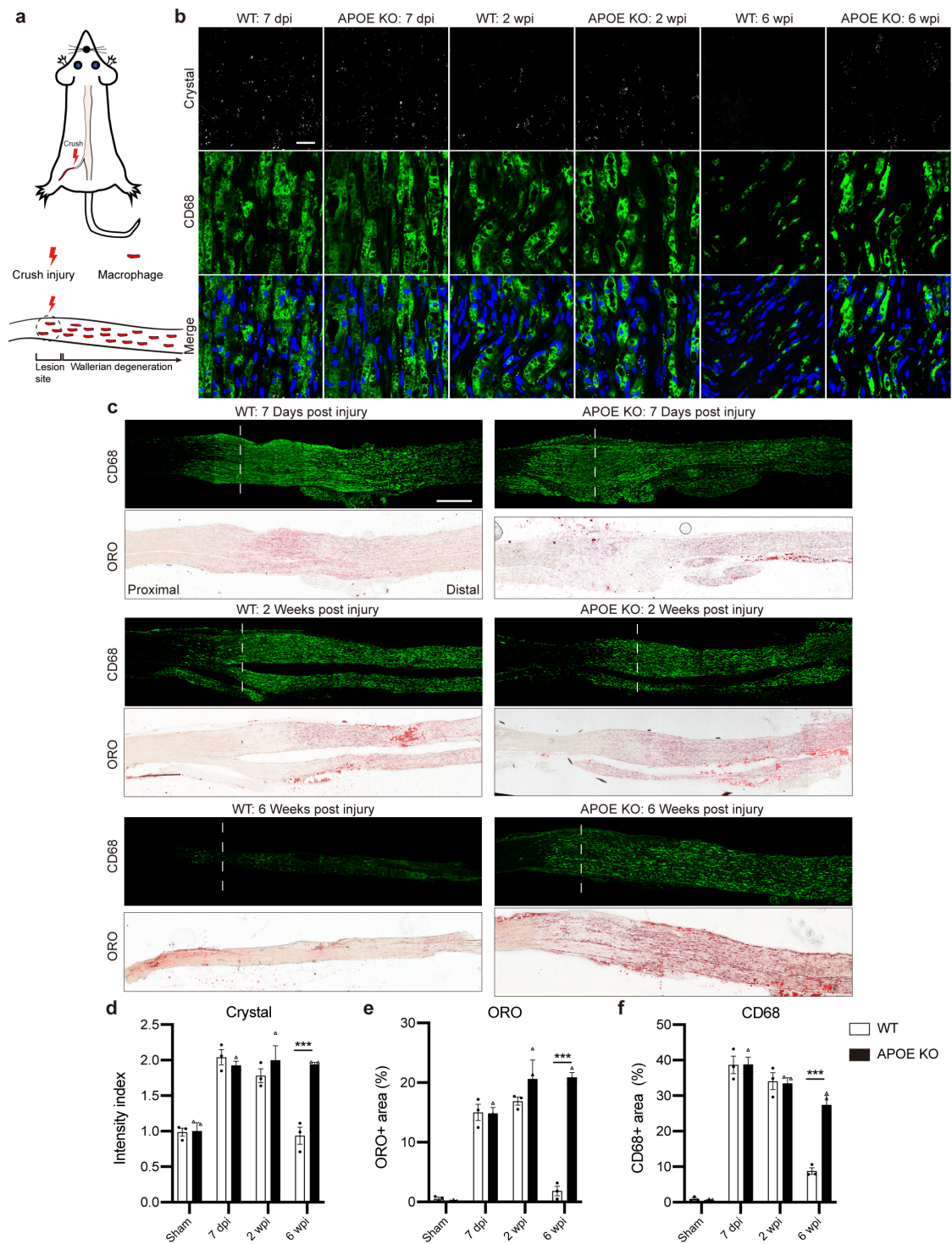

**Fig.3: Excess cholesterol accumulation is resolved by reverse cholesterol transport in injured sciatic nerves. a**, Schematic of sciatic nerve crush injury. **b**, Representative images of sciatic nerve distal to the crush site at different time points after injury, showing crystal (white), CD68 (green) and DAPI (blue).

**c**, Representative images of the sciatic nerve at different time points after injury stained with CD68 (green) and ORO (red). Dashed lines indicate the injury site. **d-f**, Quantification of crystal intensity (normalized to the proximal region), ORO positive area (%) and CD68 positive area (%) in sciatic nerve distal to the injury site (n=3 mice), Ordinary two-way ANOVA with turkey's multiple comparisons test. All data are shown as mean ± sem. ***$p < 0.001$. Scale bar: 20 μm (**b**), 500 μm (**c**).

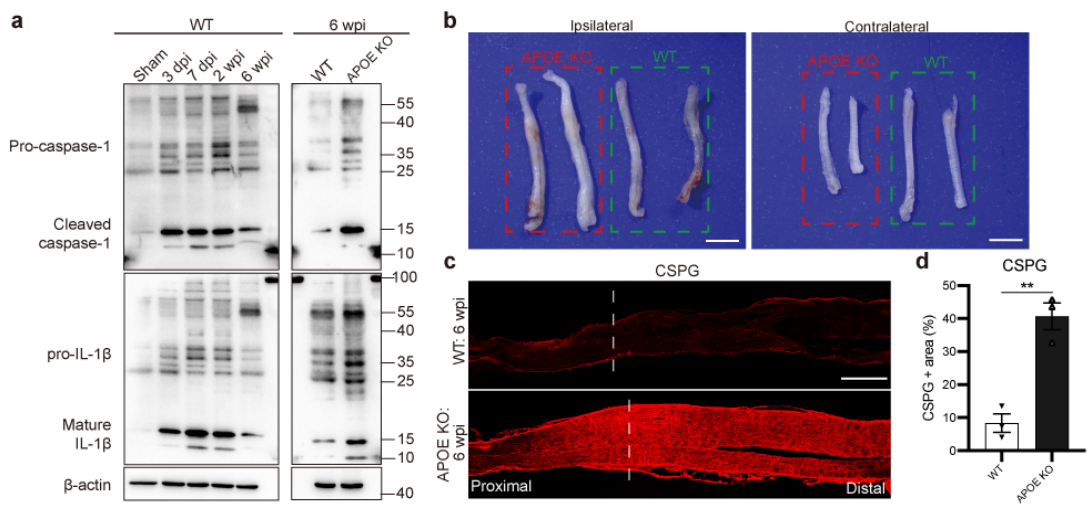

**Fig.4: Efficient cholesterol clearance prevents NLRP3 inflammasome activation and CSPG scaring. a**, Immunoblot of sciatic nerves distal to the injury site for caspase-1, IL-1β and β-actin (loading control). **b**, Images of ipsilateral and contralateral sciatic nerves of APOE KO mice and WT mice at 6 wpi. **c**, Representative images of the sciatic nerve stained with CSPG at 6 wpi. Dashed lines indicate the injury site. **d**, Quantification of CSPG positive area (%) in sciatic nerve distal to the injury site (n=3 mice). Two-tailed Student's t

test. Data are shown as mean ± sem. **p < 0.01. Scale bar: 500 μm (**d**), 2 mm (**b**).

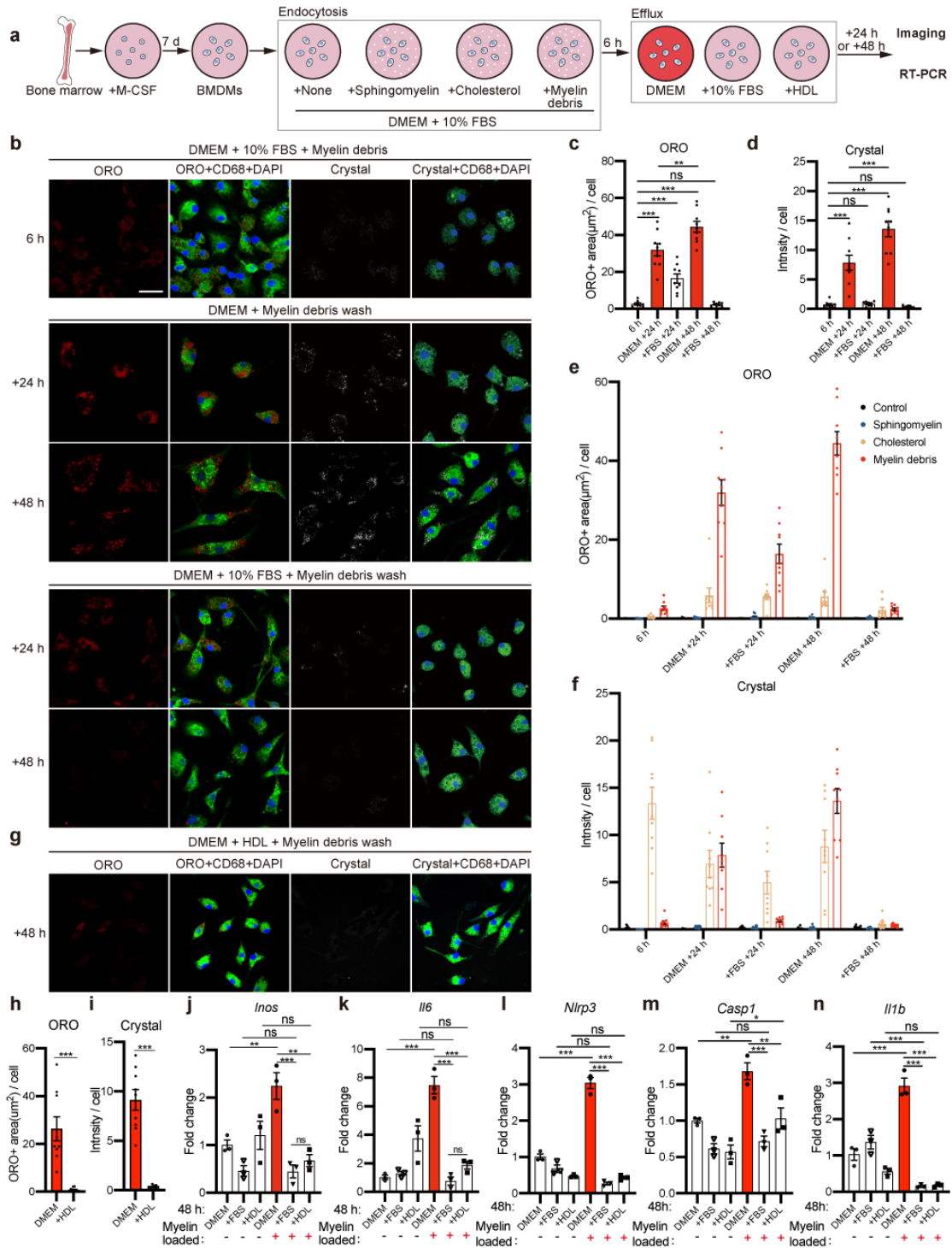

**Fig.5: Serum is efficient to remove myelin-derived cholesterol from macrophages *in vitro*. a**, Scheme of cholesterol efflux study *in vitro*. **b**, Representative images of myelin-overloaded BMDMs in different mediums at different time points with ORO (red), CD68 (green) and Crystal (white). **c**, **d**, Quantification of ORO positive area per cell and crystal intensity per cell after myelin treatment at different time points. **e**, **f**, Quantification of ORO positive area per cell and crystal intensity per cell after different treatments at different time points. The data of myelin treatment are also shown in **c**, **d** (n=9 from 3 cultures). **g**, Representative images of myelin-overloaded BMDMs after incubating in DMEM supplemented with HDL for 48 hours, showing nearly no deposition of ORO positive lipid droplets (red) and Crystals (white) in CD68 positive macrophages (green). **h**, **i**, Quantification of ORO positive area per cell and crystal intensity per cell from **d** (n=9 from 3 cultures). **j-n**, Quantitative PCR analysis of *Inos*, *IL6, Nlrp3, Casp1* and *Il1b* at 48 hours after myelin treatment in different mediums (n=3 cultures). **c**, **d**, **j-n**, Ordinary one-way ANOVA with turkey's multiple comparisons test. **h**, **i**, Two-tailed Student's t test. All data are shown as mean ± sem. *$p < 0.05$, **$p < 0.01$, ***$p < 0.001$, ns: not significant. Scale bar: 20 μm (**a**, **g**).

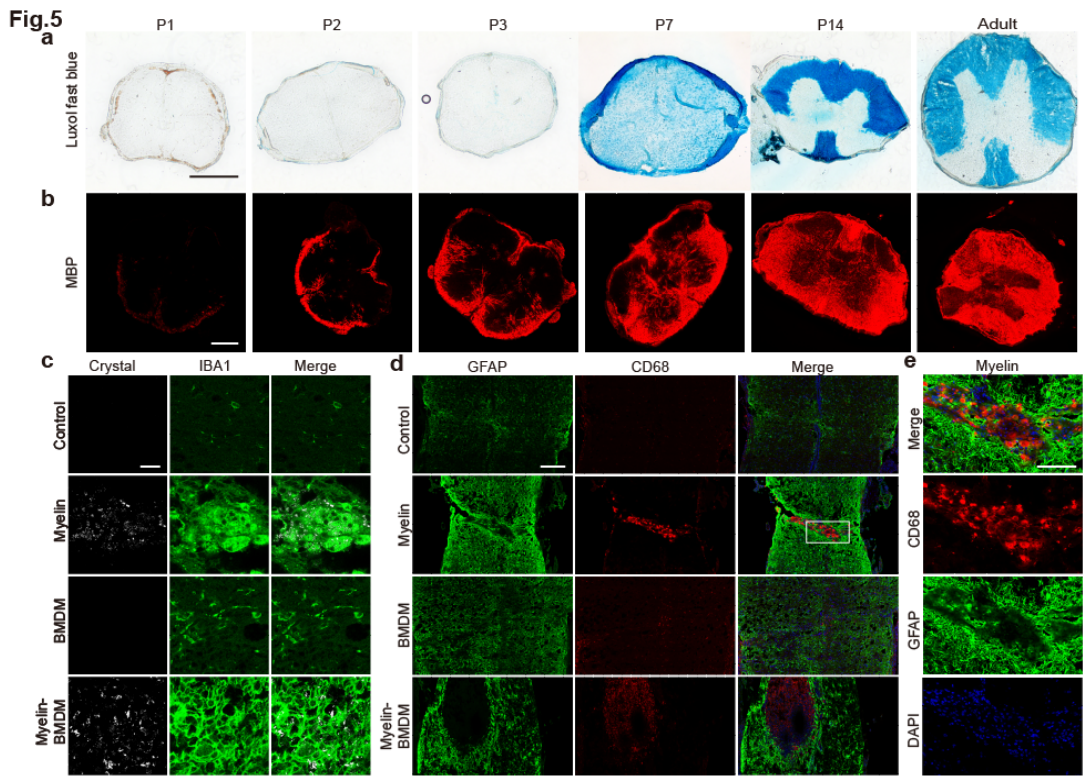

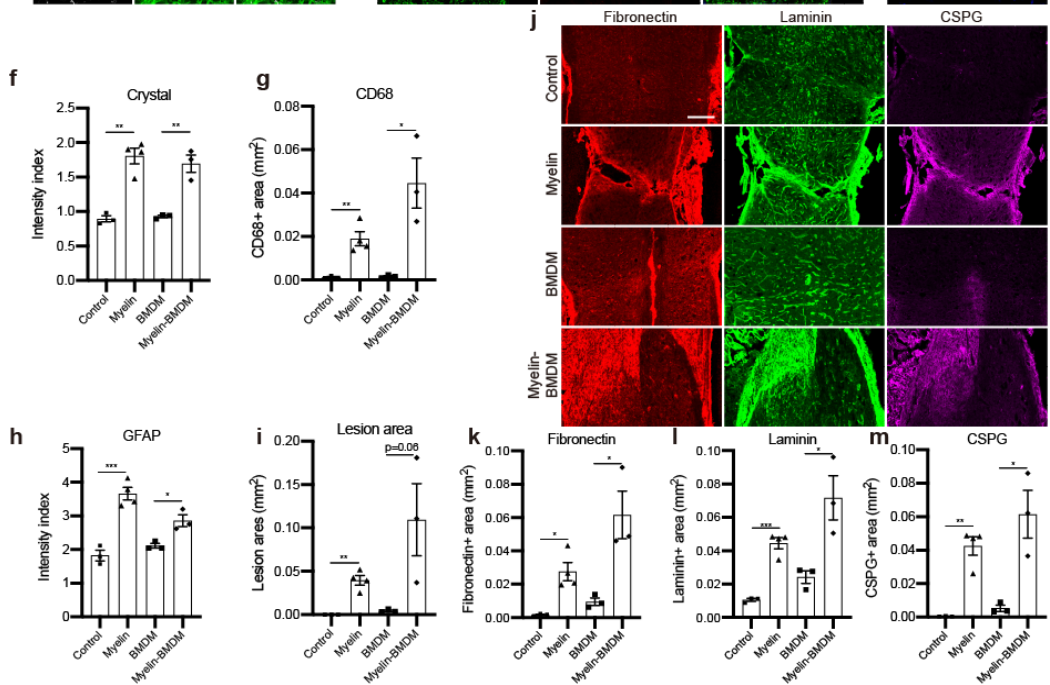

**Fig.6: Excess accumulation of myelin-derived cholesterol leads to scar-forming. a**, **b**, Representative images of postnatal and adult spinal cord stained with luxol fast blue (blue) and MBP (red), respectively. **c-e**, Representative images of spinal cord sections at 2 weeks after P2 injury with transplantation of vehicle, myelin, BMDMs or Myelin-BMDMs. **c**, Representative images showing crystal (white) and IBA1 (green). **d**, Representative images stained with GFAP (green), CD68 (red) and DAPI (blue). **e**, Higher magnification images of boxed area in **d**, showing CD68 positive GFAP negative non-neural lesion core (red) and GFAP positive glial scar (green). **f-i**, Quantification of crystal and intensity (normalized to the intact region) and CD68 positive area in the lesion, GFAP intensity (normalized to the intact region) at the lesion border at 2 wpi (n=3, 4, 3 and 3 mice for Control, Myelin, BMDMs and Myelin-BMDMs, respectively). **j**, Adjacent sections of spinal cord at 2 weeks after P2 injury with transplantation of vehicle, myelin, BMDMs or Myelin-BMDMs stained with Fibronectin (red), Laminin (green), CSPG (magenta). **k-m**, Quantification of Fibronectin, Laminin and CSPG positive area in the lesion at 2 wpi (n=3, 4, 3 and 3 mice for Control, Myelin, BMDMs and Myelin-BMDMs, respectively). **f-i**, **k-m**, Two-tailed Student's t test for control vs. myelin and BMDMs vs. Myelin-BMDMs. All data are shown as mean ± sem. $*p < 0.05$, $**p < 0.01$, $***p < 0.001$, Scale bar: 200 µm (**a**, **b**, **d**, **j**), 20 µm (**c**), 100 µm (**e**).

**Supplementary Figures**

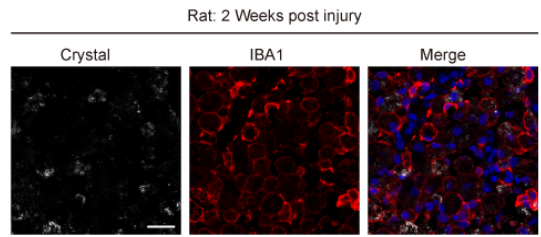

**Supplementary Fig.1: Cholesterol crystals appear in rat spinal cord lesion.**

Representative images of rat spinal cord lesions at 2 wpi, showing crystals (white) in IBA1 positive phagocytes (red). Scale bar: 20 μm.

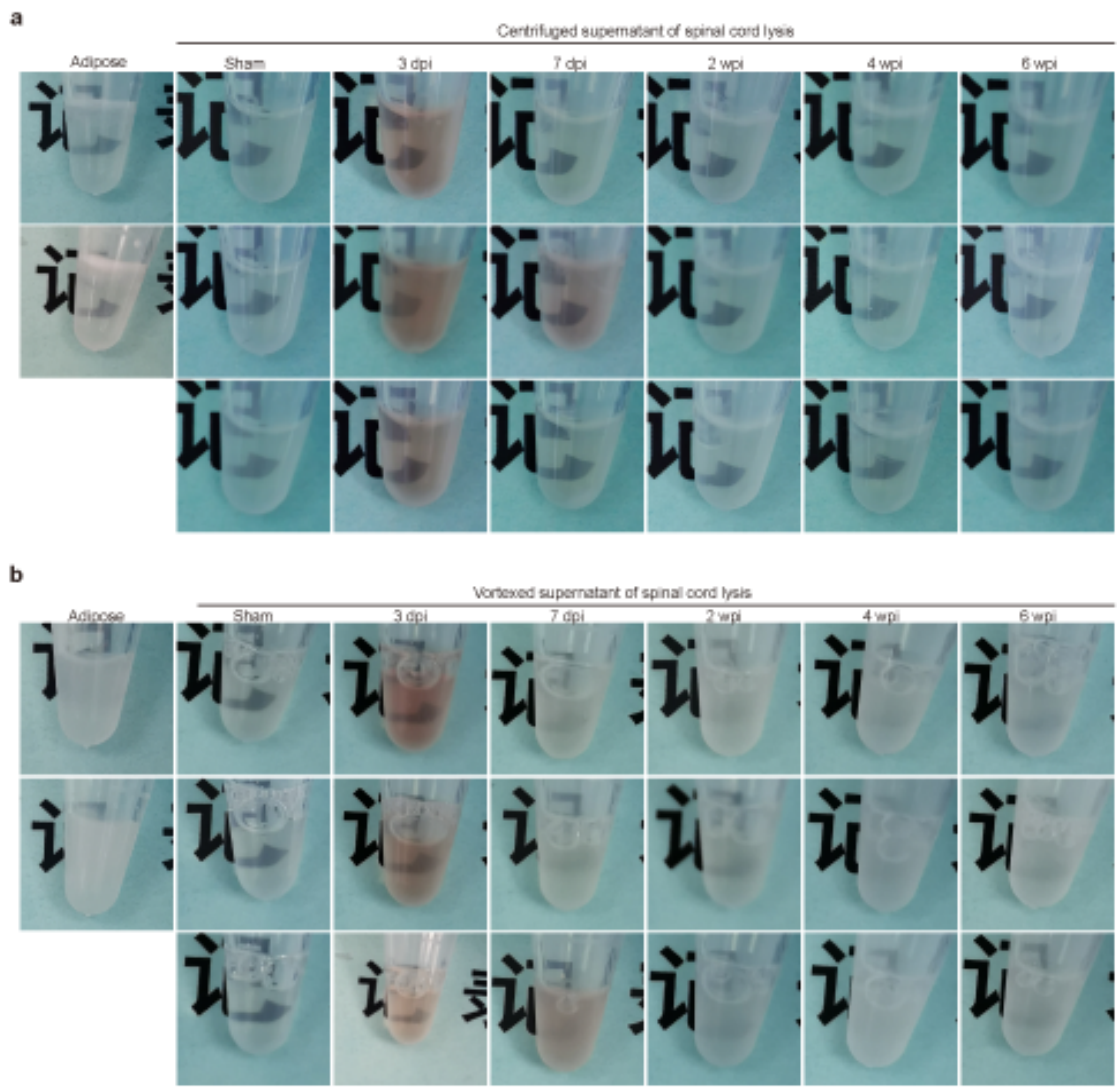

**Supplementary Fig.2: A Lipid layer appears on the supernatant of spinal cord lysate after injury. a**, Images of centrifuged supernatants of spinal cord lysates and adipose lysates (positive control), showing a white layer on the top of supernatants from 7 dpi to 6 wpi, but it is not easy to distinguish in images. **b**, Images of vortexed supernatants of spinal cord lysates and adipose lysates, showing the supernatants of the spinal cord were opaque from 7 dpi to 6 wpi, similar to the supernatants of adipose lysates.

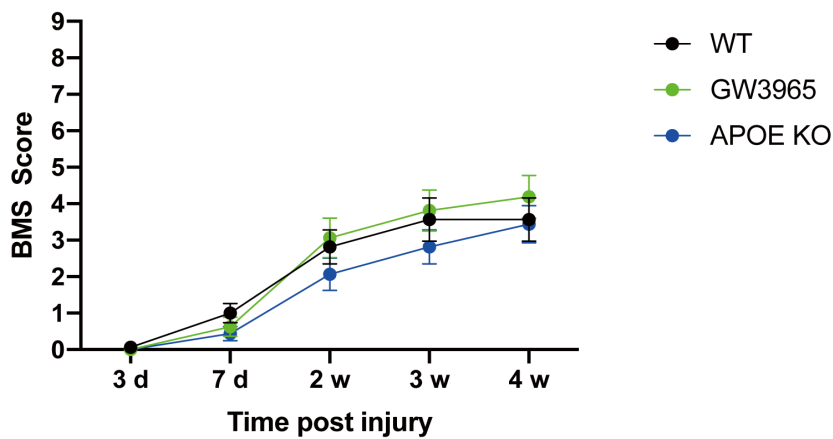

**Supplementary Fig.3: Locomotion recovery evaluation.** Locomotor recovery experiments were evaluated with the BMS scores (n=8 mice). No significant differences were observed among any of the experimental groups. Repeated measures two-way ANOVA with the Geisser-Greenhouse correction. Data are shown as mean ± sem.

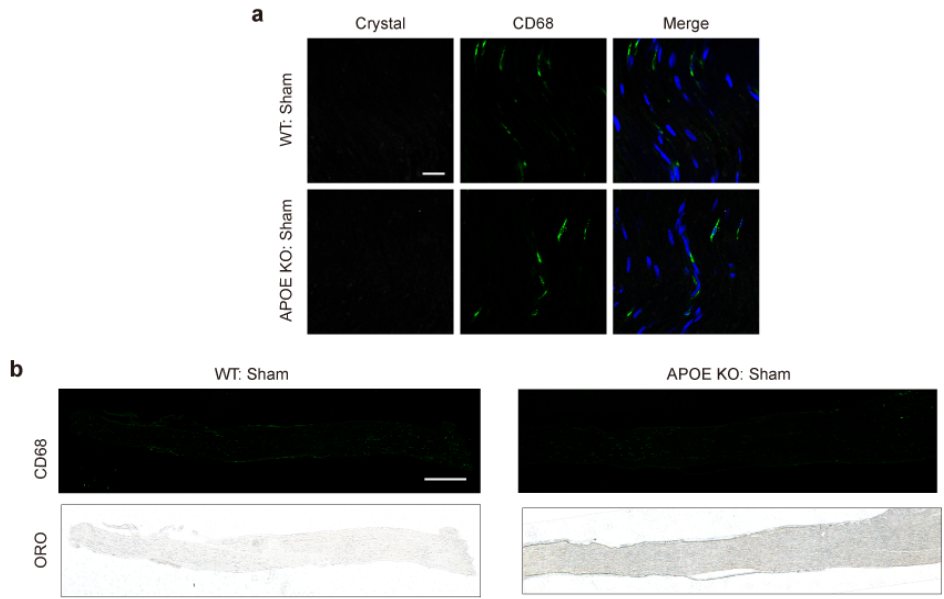

**Supplementary Fig.4: No cholesterol deposition in uninured sciatic nerves. a**, Representative images of uninjured sciatic nerve of WT and APOE KO mice, showing crystal (white), CD68 (green) and DAPI (blue). **b**, Representative images of uninjured sciatic nerve of WT and APOE KO mice stained with CD68 (green) and ORO (red). Scale bar: 20 μm (**a**), 500 μm (**b**).

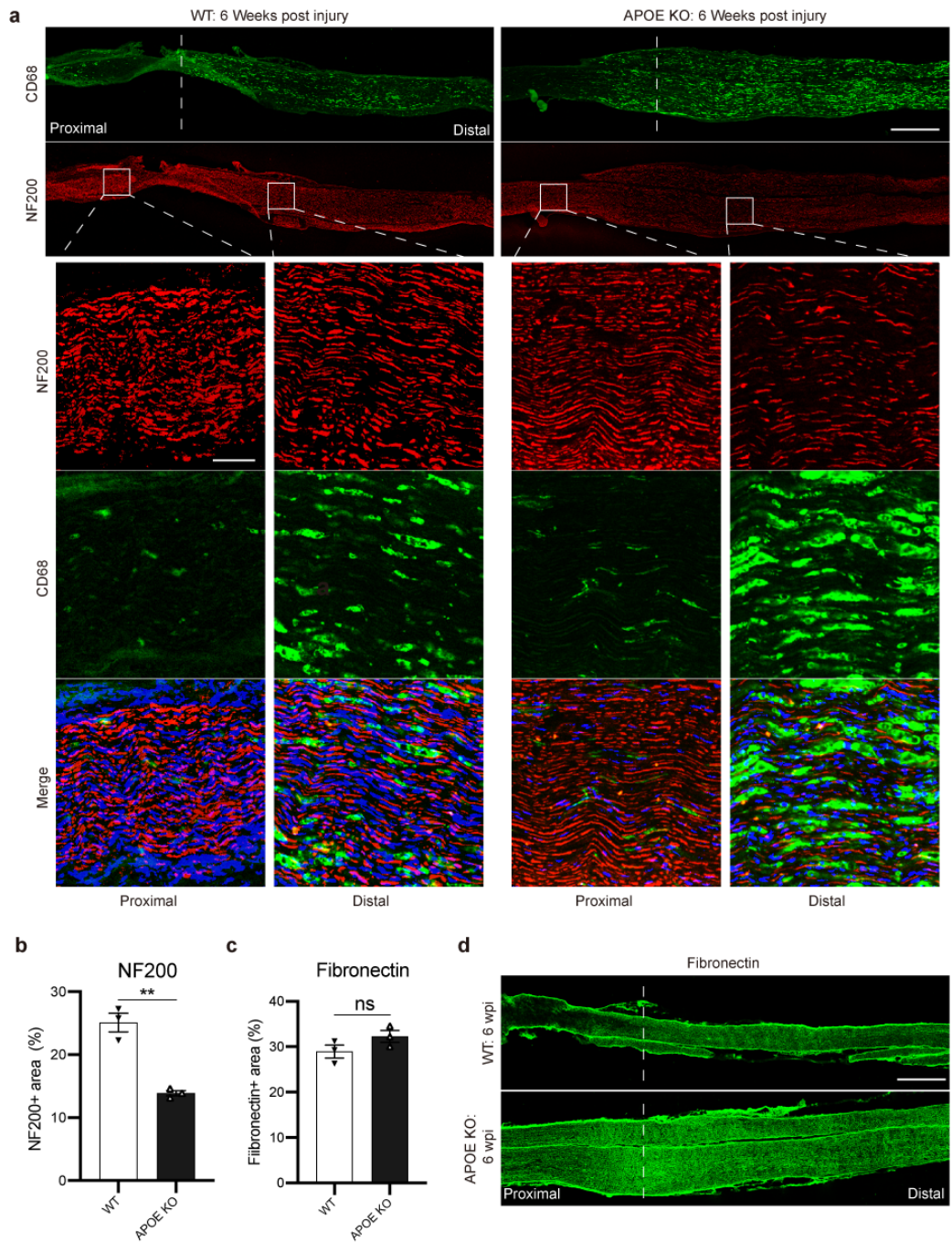

**Supplementary Fig.5: Limited cholesterol clearance reduces nerve fiber density. a**, Representative images of the sciatic nerve from APOE KO and WT mice at 6 wpi stained with CD68 (green), NF200 (red) and DAPI (blue). Higher magnification images of boxed areas show the density of nerve fiber (red) and CD68 positive macrophages (green). Dashed lines indicate the injury site. **b**,

Quantification of NF200 positive area in sciatic nerve distal to the injury site at 6 wpi (n=3 mice). **c**, **d**, Representative images of the sciatic nerve stained with fibronectin (green) and quantification of fibronectin positive area (%) in sciatic nerve distal to the injury site at 6 wpi (n=3 mice). **b**, **c**, Two-tailed Student's t test. All data are shown as mean ± sem. ***p < 0.001, ns: not significant. Scale bar: 500 µm (**a**, **d**), 50 µm (**a**, higher magnification images).

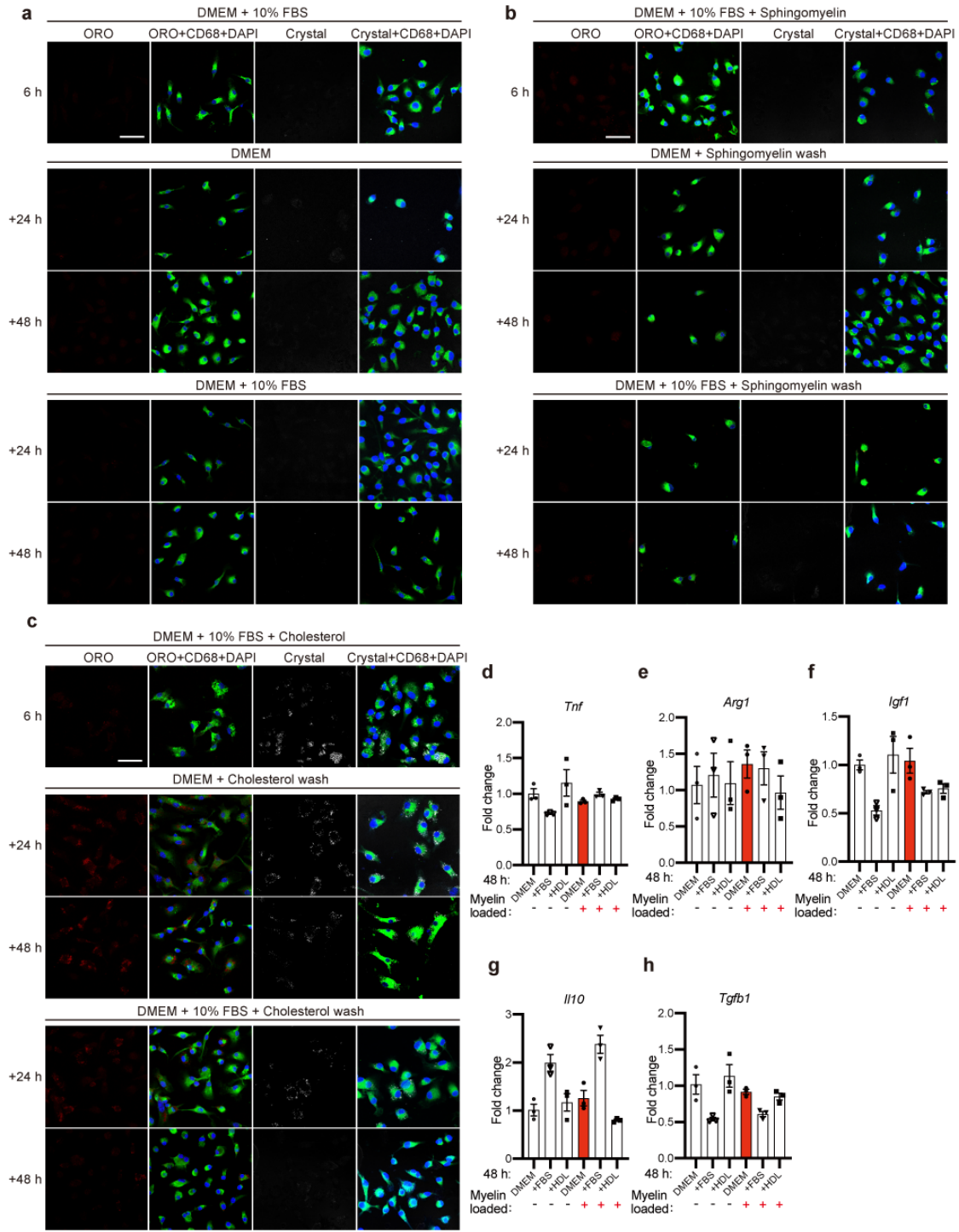

**Supplementary Fig.6: Deposition and removal of cholesterol from macrophages *in vitro*. a-c**, Representative images of BMDMs after indicated treatments at different time points showing ORO (red), CD68 (green), Crystal (white) and DAPI (blue). Scale bar: 20 µm. **d-h**, Real-time PCR analysis of *Arg1,*

*Tnf*, *Igf1*, *Il10* and *Tgfb1* in myelin-overloaded BMDMs after incubating in different mediums for 48 hours (n=3 cultures). Ordinary one-way ANOVA with turkey's multiple comparisons test. Data are shown as mean ± sem.

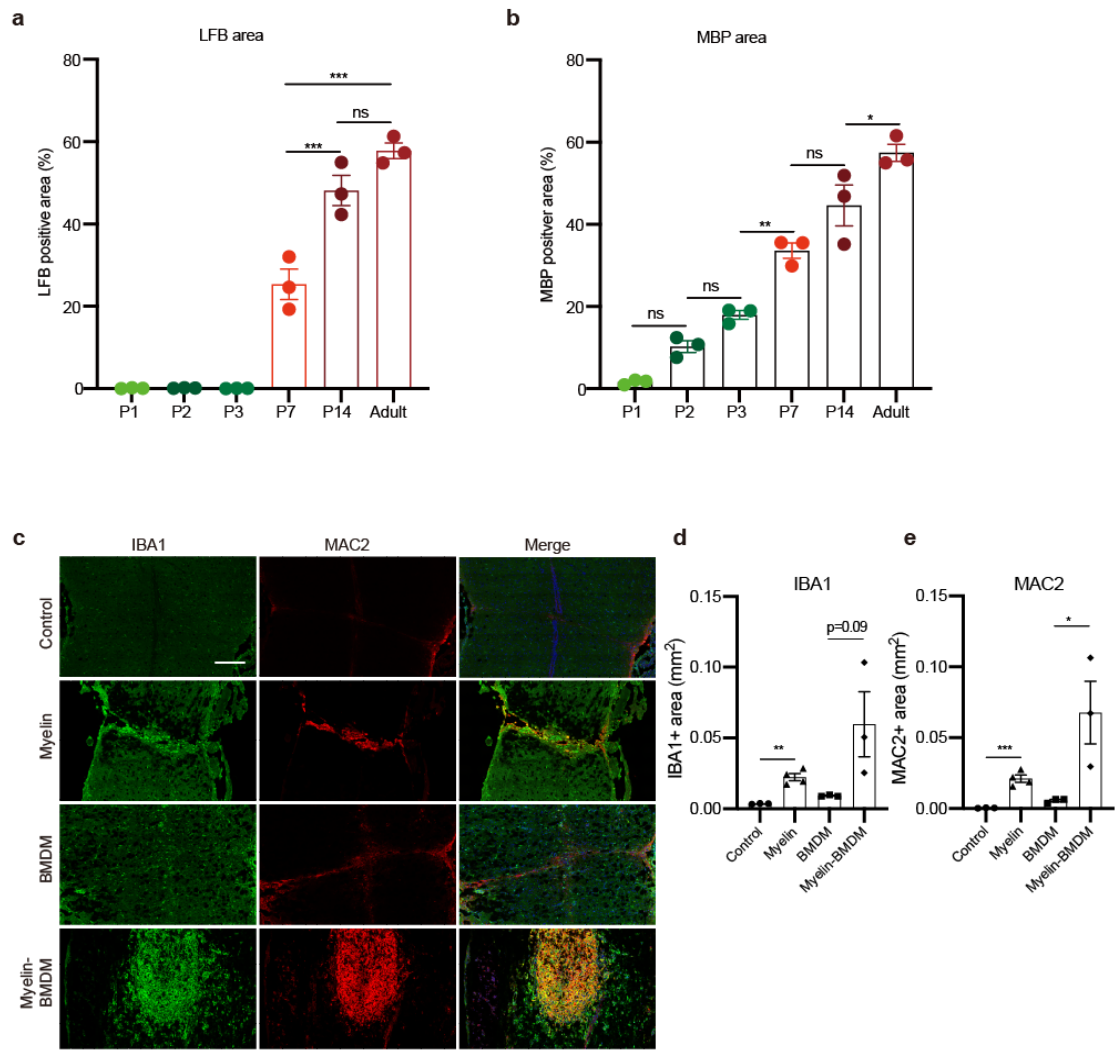

**Supplementary Fig.7: Deposition of macrophages in spinal cord lesions on account of lipids derived from myelin. a**, **b**, Quantification of LFB and MBP positive area (%) in sections of the postnatal and adult spinal cord, respectively (n=3 mice). **c**, Representative images of spinal cord lesion at 2 wpi

stained with IBA1 (green), MAC2 (red) and DAPI (blue). Scale bar: 200 μm. **d**, **e**, Quantification of IBA1 and MAC2 positive area in spinal cord lesion at 2 wpi (n= 3, 4, 3 and 3 mice for Control, Myelin, BMDM and Myelin-BMDM, respectively). **a**, **b**, Ordinary one-way ANOVA with turkey's multiple comparisons test. **d**, **e**, Two-tailed Student's t test for control vs. myelin and BMDMs vs. Myelin-BMDMs. All data are shown as mean ± sem. $*p < 0.05$, $**p < 0.01$, $***p < 0.001$.

**Supplementary Table 1: List of primer sequences for real-time PCR.**

| Gene | Forward primer (5'-3') | Reverse primer (5'-3') |
| --- | --- | --- |
| *Hmgcr* | GCTCGTCTACAGAAACTCCACG | GCTTCAGCAGTGCTTTCTCCGT |
| *Fdft1* | GGATGTGACCTCCAAACAGGAC | CAGACCCATTGAGTTGGCACAC |
| *Cyp51* | ATCCAGAAGCGCAGGCTGTCAA | CAGTCCGATGAGCATCCCTGAT |
| *Dhcr24* | CTGGAGAACCACTTCGTGGAAG | CTCCACATGCTTGAAGAACCAGG |
| *Sqle* | TGTTGCGGATGGACTCTTCC | GTTGACCAGAACAAGCTCCGCA |
| *Srebf2* | AGAAAGAGCGGTGGAGTCCTTG | GAACTGCTGGAGAATGGTGAGG |
| *Cd36* | GGAGCCATCTTTGAGCCTTCA | GAACCAAACTGAGGAATGGATCT |
| *Trem2* | CTACCAGTGTCAGAGTCTCCGA | CCTCGAAACTCGATGACTCCTC |
| *Ldlr* | GAATCTACTGGTCCGACCTGTC | CTGTCCAGTAGATGTTGCGGTG |
| *Vldlr* | ACGGCAGCGATGAGGTCAACTG | CAGAGCCATCAACACAGTCTCG |
| *Apobr* | GGATGTTCACAGCAACTGGAATG | GTCACACTGTGGCTCAGGAACA |
| *Lrp1* | CGAGAGCCTTTGTGCTGGATGA | CGGATGTCCTTCTCAATGAGGG |
| *Npc1* | TGAGGTCATCCCATTCCTGGTG | TCCAGCGTTTCCTCCTGAAGAC |
| *Npc2* | GCCAGTCCTACAGTGTCAACATC | TCTTACAACCGTCAGGCTCAGG |
| *Tspo* | GAGCCTACTTTGTACGTGGCGA | GCTCTTTCCAGACTATGTAGGAG |
| *Acat1* | TGAGAGCACCTCCAGAACAAGG | GGACGAATAGGATGAGGAGTGC |
| *Nceh1* | CGGTATTTCTGGAGACAGTGCTG | GGTGTGTTGAAGTCCAAAGCCTG |
| *Cyp46a1* | CTCAGGACGATGAGGTTCTGCT | TGGCGAGACAACTCCATCACTG |
| *Ch25h* | CCTAAGTCACGTCCTGATCTGC | GAGGACGAGTTCTGGTGATGCA |
| *Cyp27a1* | ATGGGATCTTCATCGCACA | CGTTTAAGGCATCCGTGTAGA |
| *Abca1* | TGTCTGAAAAAGGAGGACAGTG | TGTCACTTTCATGGTCGCTG |
| *Abcg1* | CAGACGAGAGATGGTCAAAGA | TCAAAGAACATGACAGGCGG |
| *Apoe* | CTGACAGGATGCCTAGCC | TCCCAGGGTTGGTTGCTTTG |
| *Apoj* | GATGATCCACCAGGCTCAACAG | ACACAGTGCGGTCATCTTCACC |
| *Apod* | GGTGAAGCCAAACAGAGCAACG | CAGGAGTACACGAGGGCATAGT |
| *Pltp* | GCTGCTGAACATCTCCAACGCA | GCTGTAGACCTGTTCGGATGGA |
| *Lcat* | CAGTCCTGGAAGGACCACTTCA | GAAGTCGTGGTTATGCGCTGCT |
| *Il6* | TACCACTTCACAAGTCGGAGGC | CTGCAAGTGCATCATCGTTGTTC |
| *Inos* | GAGACAGGGAAGTCTGAAGCAC | CCAGCAGTAGTTGCTCCTCTTC |
| *Tnf* | TCTTCTCATTCCTGCTTGTGG | GGTCTGGGCCATAGAACTGA |
| *Arg1* | CATTGGCTTGCGAGACGTAGAC | GCTGAAGGTCTCTTCCATCACC |
| *Il10* | CGGGAAGACAATAACTGCACCC | CGGTTAGCAGTATGTTGTCCAGC |
| *Igf1* | GTGGATGCTCTTCAGTTCGTGTG | TCCAGTCTCCTCAGATCACAGC |
| *Tgfb1* | TGATACGCCTGAGTGGCTGTCT | CACAAGAGCAGTGAGCGCTGAA |
| *Nlrp3* | TCACAACTCGCCCAAGGAGGAA | AAGAGACCACGGCAGAAGCTAG |
| *Casp1* | GGCACATTTCCAGGACTGACTG | GCAAGACGTGTACGAGTGGTTG |
| *Il1b* | GCAACTGTTCCTGAACTCAACT | ATCTTTTGGGGTCCGTCAACT |
| *Actb* | GCTTCTAGGCGGACTGTTACTGA | GCCATGCCAATGTTGTCTCTTAT |